# Quantized decay charges in non-Hermitian networks characterized by directed graphs

Wenwen Liu[1, #], Junyao Wu[2, #], Li Zhang[1, 2, #], Oubo You[1], Ye Tian[5], Hongsheng Chen[2], Bumki Min[3†], Yihao Yang[2*], Shuang Zhang[1,4,6,7§]

[1] *New Cornerstone Science Laboratory, Department of Physics, The University of Hong Kong, Hong Kong, 999077, China.*
[2] *Interdisciplinary Center for Quantum Information,*
*State Key Laboratory of Modern Optical Instrumentation, Zhejiang University, Hangzhou 310027, China.*
[3] *Department of Physics, Korea Advanced Institute of Science and Technology, Daejeon 34141, Republic of Korea.*
[4] *Department of Electrical & Electronic Engineering, The University of Hong Kong, Hong Kong 999077, China.*
[5] *Department of Mechanical Engineering, The University of Hong Kong, Hong Kong 999077, China.*
[6] *State Key Laboratory of Optical Quantum Materials, University of Hong Kong, Hong Kong 999077, China.*
[7] *Materials Innovation Institute for Life Sciences and Energy (MILES), HKU-SIRI, Shenzhen 518000, China.*

Non-Hermitian physics has unveiled a realm of exotic phenomena absent in Hermitian systems, with the non-Hermitian skin effect (NHSE) showcasing boundary-localized eigenstates driven by non-reciprocal interactions. Here, we introduce a new class of non-Hermitian systems exhibiting pure decay modes—eigenstates with pure, smooth exponential decay, devoid of the oscillatory wave patterns typical of traditional NHSE. Modeled as directed graphs with non-reciprocal hopping, these systems reveal quantized decay charges, defined as the sum of decay constants along edges at each node, offering a novel topological invariant. We derive universal conditions for these modes, enabling versatile configurations from one-dimensional rings, directed graphs with complicated connectivity, to higher-dimensional lattices. Experimental validation using microwave resonant circuits confirms the predicted pure decay profiles. This discovery paves the way for potential applications in photonics, signal processing, and beyond, harnessing the unique topological properties of non-Hermitian networks.

In the quest to understand the fundamental principles that govern the behavior of physical systems, physicists have long been guided by the Hermitian nature of operators in quantum mechanics. However, recent strides in research have led to the exploration of a fascinating and unconventional domain known as non-Hermitian physics [1-5]. Departing from the familiar constraints of Hermitian operators, non-Hermitian physics opens doors to a rich tapestry of phenomena, including the emergence of complex eigenvalues [6], non-reciprocal wave propagation [7, 8], and exceptional points [9-11]. Such unconventional behavior has profound implications for our understanding of quantum mechanics [5, 12], acoustic systems [13, 14], and in the realm of optics and photonics [15, 16].

The non-Hermitian skin effect (NHSE), a captivating manifestation of non-Hermitian physics originating from nonreciprocal hopping, describes phenomenon that a large number of eigenstates become localized at the boundaries of the system [17-24] and has the potential to redefine the landscape of unidirectional amplifiers [25], energy harvesting [26], and signal processing technologies [27, 28]. Despite its profound implications, the wavefunctions are not perfectly decaying in an exponential way. Instead, the states are generally forming standing wave patterns with strong oscillatory features.

In this work, we introduce a new class of non-Hermitian systems that support pure decay modes—eigenstates characterized by smooth, non-oscillatory exponential decay. Represented as directed graphs with non-reciprocal hopping, these systems exhibit a striking feature: quantized decay charges, defined as the sum of decay constants along edges at each node. We establish the conditions for these modes, demonstrating their existence in diverse configurations, from one-dimensional (1D) rings, directed graphs of complex configuration, to higher-dimensional lattices. Our theoretical framework, validated through microwave resonator experiments, unveils a new topological paradigm with potential applications in photonic and acoustic lattices, where precise control of directional decay is paramount.

We start with the simplest configuration that exhibits pure decay modes - a loop comprising two joined sections named as type-A chain (site 1 to $N$) and type-B chain (site $N+1$ to $N+M$) with opposite hopping directions, as depicted in Fig. 1(a). With fixed hopping coefficients between adjacent sites, only the hopping direction matters, enabling us to model the loop as a directed ring. Thus, we use $t = t_r/t_l$ to represent the normalized non-reciprocal hopping parameter, and the corresponding model is described as:

$$H = \sum_{i=1}^{N>1} \left(t c_i^\dagger c_{i+1} + c_{i+1}^\dagger c_i\right) + c_{N+M}^\dagger c_1 + t c_1^\dagger c_{N+M} + \sum_{i=\mathrm{N}+1}^{N+M-1,\, M>1} \left(c_i^\dagger c_{i+1} + t c_{i+1}^\dagger c_i\right) \qquad (1)$$

Onsite energy is uniform across all nodes and thus neglected. The two boundary sites, colored in red (drain) and blue (source), divide the loop into the two opposite hopping chains. For $N = 29$ and $M = 1$, the wave functions for all the eigenmodes are shown in Fig. 1(b), exhibiting identical amplitude distribution. All the eigenstates are localized at the 30th site, displaying purely exponential wavefunctions with decay constants of $t^{-\frac{1}{30}}$ and $t^{-\frac{29}{30}}$ along the two opposite directions. When $N$ and $M$ are varied, the location of the localized state shifts while the pure exponential profile persists, as shown in Fig. 1(c) with $[N, M] = [13, 17]$. For comparison, we also examine the traditional NHSE with open boundary condition (OBC), i.e., $H = \sum_{i=1}^{N}(c_i^{\dagger}c_{i+1} + tc_{i+1}^{\dagger}c_i)$, as shown in Fig. 1(d). The $n_{th}$ eigen-state has a wave function of $\psi_m = r_0^m \sin(m\theta_n)$ on the $m_{th}$ site, where $r_0 = t^{-\frac{1}{2}}$ and $\theta_n = \frac{\pi n}{N+1}$. This typically exhibits standing wave-like oscillations along sides decay, in sharp contrast to the purely exponential profile of our configurations.

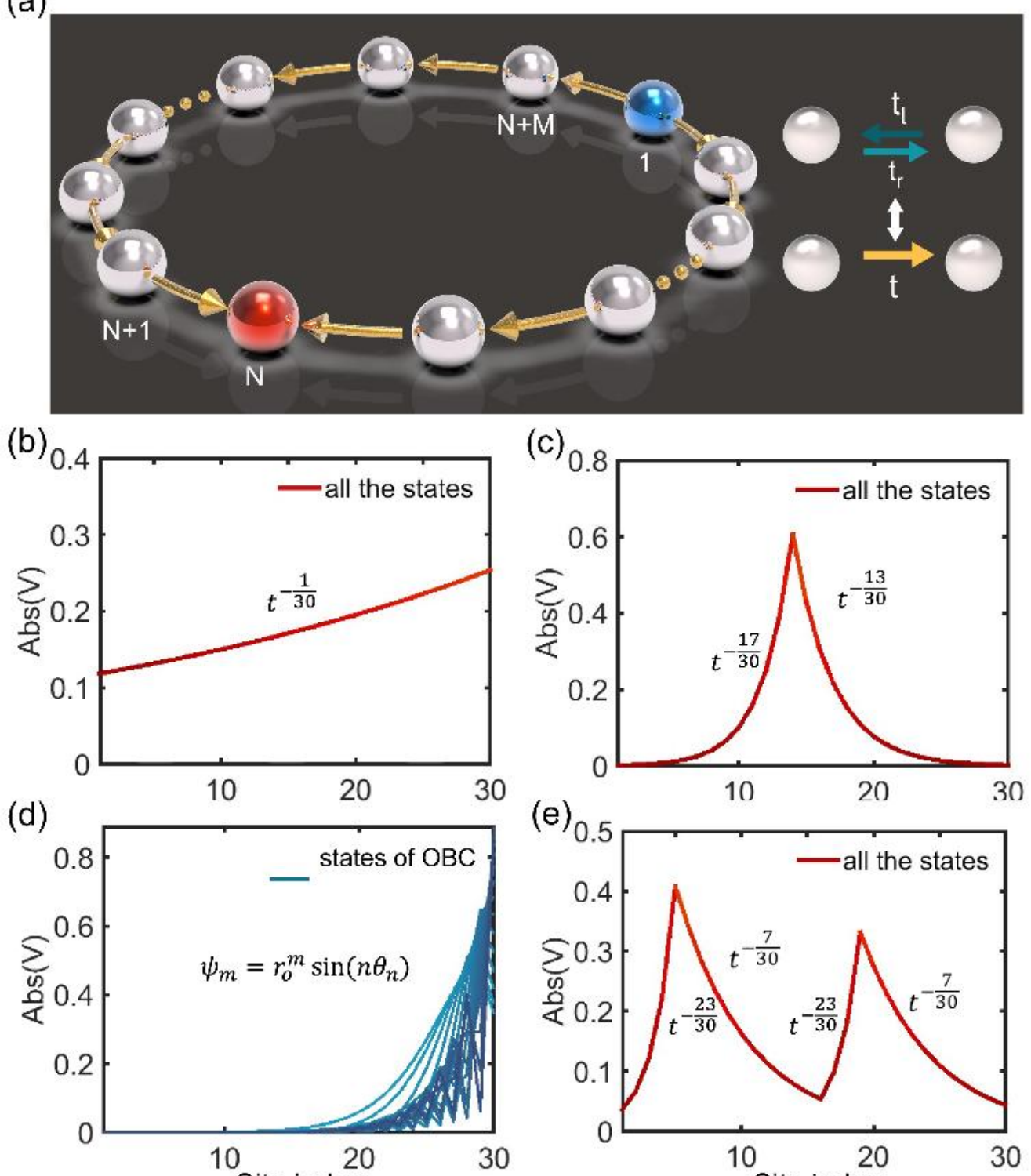

**Fig.1 Directed rings that support pure decay modes.** (a) Schematic view of the general hopping model that have two chains holding opposite hopping directions connecting together. The bulk sites are colored light gray, the boundaries are colored red and blue, respectively. The right inset shows the equivalence between two representations of the non-reciprocal hopping. We use $t > 1$ for illustration through the whole paper. (b-c) Amplitude of wavefunction distribution when $[N, M] = [29, 1]$ (b) and when $[N, M] = [13, 17]$ (c). (d) Amplitude of wavefunction distribution along each site under OBC. (e) Amplitude distribution of two type-A chains and two type-B chains linked together with $[N_1, M_1, N_2, M_2] = [4, 11, 3, 12]$.

To understand the pure decay phenomenon, we analyze the directed ring using the generalized Brillouin-zone method. We start with the simple case of a loop consisting of two chain sections of opposite hopping directions (one type-A chain and one type-B chain), while more general configurations consisting of multiple sections are given in the Supplemental Material [29]. The loop is described by two bulk equations:

$$\begin{aligned} t\varphi_{m-1} - E\varphi_m + \varphi_{m+1} = 0 \\ \psi_{m-1} - E\psi_m + t\psi_{m+1} = 0 \end{aligned} \tag{2}$$

To solve for the equation, we take the single wave vector ansatz of the wave function, $\varphi_m \propto \alpha_j^m$ ($\psi_m \propto \beta_j^m$) for type-A (type-B) chain and the bulk equations are transformed to $t\alpha_j^{m-1} - E\alpha_j^m + \alpha_j^{m+1} = 0$ ($\beta_j^{m-1} - E\beta_j^m + t\beta_j^{m+1} = 0$). For given $E$, there are two solutions of the ansatz, denoted as $\alpha_{1,2}$ and $\beta_{1,2}$, which forms general solutions of the bulk equations: $\varphi_m = C_1{\alpha_1}^m + C_2\,{\alpha_2}^m$ for type-A and $\psi_m = C_3{\beta_1}^m + C_4{\beta_2}^m$ for type-B. Applying the boundary continuity, we obtain $M[C_1, C_2, C_3, C_4]^T = 0$, where

$$M = \begin{bmatrix} {\alpha_1}^{N+1} & {\alpha_2}^{N+1} & -\beta_1 & -\beta_2 \\ \alpha_1 & \alpha_2 & -{\beta_1}^{M+1} & -{\beta_2}^{M+1} \\ \frac{1}{t^2}\alpha_1^2 & \frac{1}{t^2}\alpha_2^2 & -{\beta_1}^{M+2} & -{\beta_2}^{M+2} \\ \frac{1}{t^{N+2}}\alpha_1^{N+2} & \frac{1}{t^{N+2}}\alpha_2^{N+2} & -{\beta_1}^2 & -{\beta_2}^2 \end{bmatrix} \tag{3}$$

By solving $det[M] = 0$, we get $\alpha_1 = t^{\frac{M}{M+N}}e^{ik_r}$, where $k_r = \frac{2\pi n}{M+N}$ is the phase of the $n_{th}$ mode. Substituting $\alpha_1$ back to Eq. (3) reveals that the first and fourth column of the matrix are linearly dependent, resulting in $C_1 = C_4 = 0$ (or $C_2 = C_3 = 0$), such that $\varphi_n$ and $\psi_n$ become exponential functions [29]. Interestingly, the decay constants in type-A and type-B chains, given by $D = |\frac{\varphi_m}{\varphi_{m+1}}| = t^{-\frac{M}{N+M}}$ and $G = |\frac{\psi_{m+1}}{\psi_m}| = t^{-\frac{N}{N+M}}$, exhibit a simple relation $|\log_t D| + |\log_t G| = 1$. This sum remains constant when $N$ and $M$ are varied, indicating that the system follows a simple power partition rule. Such power partition rule also applies to loops consisting of arbitrary number of sections of directed chains alternating between type-A and type-B, with a detailed derivation provided in the Supplemental Materials [29]. For a system with $p$ type-A (type-B) chains having $[n_1, \dots, n_p]$ ($[m_1, \dots, m_p]$) sites, the corresponding decay constant is $t^{-\frac{m_1+\cdots+m_p}{n_1+m_1+\cdots+n_p+m_p}}$ ($t^{-\frac{n_1+\cdots+n_p}{n_1+m_1+\cdots+n_p+m_p}}$) for all the type-A (type-

B) chains, meaning the amplitude of the wavefunction decays with the same rate within chains of the same type. For example, Fig. 1(e) illustrates the amplitude distribution for the configuration with two type-A (B) chains, where the sum of the decay constants at each A-B chain boundary equals to 1, and the decay constants are only determined by the total number of sites in all chains of type-A and that of type-B.

Significantly, the non-Hermitian pure decay modes can go beyond loop configuration and be extended to complex directed graphs, as exemplified in Fig 2(a-b). The corresponding wave function, shown on the right, exhibits perfect exponential decay for all modes across sites without any oscillatory behavior. We now formulate the general configuration of directed graph that supports this pure decay modes. Nodes are labeled from 1 to *N*, and the graph is designed such that all the eigen-modes decay geometrically from site 1 to site *N*. The desired eigenvectors can be organized in the following normalized matrix,

$$\boldsymbol{\psi}=\begin{bmatrix} 1 & 1 & \cdots & 1 \\ r & re^{i\theta} & \cdots & re^{i(N-1)\theta} \\ \cdots & \cdots & \cdots & \cdots \\ r^{N-1} & r^{N-1}e^{i(N-1)\theta} & \cdots & r^{N-1}e^{i(N-1)(N-1)\theta} \end{bmatrix} \tag{4}$$

where $\psi_{mn}=r^{m-1}e^{i(n-1)(m-1)\theta}$ ( $r=t^{-\frac{1}{N}}$ for simplification) represents the $n_{th}$ eigen-state on the $m_{th}$ site, and $\theta=\frac{2\pi}{N}$. Notably, the matrix is a Vandermonde matrix and its inverse can be computed straightforwardly [29]. Therefore, by using the relation $\boldsymbol{H}=\boldsymbol{\psi}\,\boldsymbol{E}\,\boldsymbol{\psi}^{-1}$, where $diag(\boldsymbol{E})=[E_1,E_2,\dots,E_N]$, one can get

$$H_{a,b}=\frac{1}{N}\frac{r^{a-1}}{r^{b-1}}\sum_{n=1}^{N}E_n e^{i(a-b)(n-1)\theta} \tag{5}$$

For node 1 with all directional arrows pointing into it via in-ward hopping amplitude $t$, the energy can be written as $\psi_{1n}E_n=a_1\psi_{2n}t+\cdots+a_{N-1}\psi_{Nn}t$ , where $a_i\in\{0,1\}$ represents the connectivity between node 1 and all other node. Therefore, we have,

$$H_{1,b}=\frac{1}{N}\frac{1}{r^{b-1}}\sum_{q=1}^{N-1}\sum_{n=1}^{N}a_q t^{\frac{N-q}{N}}e^{i(1-b+q)(n-1)\theta} \tag{6}$$

It is straightforward to show that $H_{1,b}=r^{1-b}t^{\frac{N-q_0}{N}}$ with $a_{q_0}=1$ if $1-b+q_0=0$ or $N$, and $H_{1,b}=0$ otherwise. To satisfy $\frac{H_{1,b}}{H_{b,1}}=t$, we require $H_{b,1}=r^{1-b}t^{-\frac{q_0}{N}}$, meaning that $b-1+q_0'=N$, with $q_0'=N-q_0$. Thus, $a_{N-q_0}=1$ is necessary to ensure $H_{b,1}\neq 0$. This analysis shows that if node 1 receives a directional hopping from node $q+1$, it must also receive equivalent hopping from node $N-q+1$. Since other nodes have the same energy as node 1, the hopping configuration of all other nodes can be determined sequentially from Node 1's information. To this end, we obtain the general form of the Hamiltonian as

$$\boldsymbol{H}=\begin{bmatrix} 0 & a_1 t & \cdots & a_{N-1}t \\ a_1 & & & \cdots \\ & & \ddots & a_1 t \\ \cdots & & & \\ a_{N-1} & \cdots & a_1 & 0 \end{bmatrix} \tag{7}$$

As long as $a_i=a_{N-i}$ $(a_i\in\{0,1\})$ is satisfied, the pure decay mode could emerge. Such form allows us to generate the pure decay phenomenon in very rich configurations.

The directed graphs in higher dimensions are also explored, constructed through the direct product of multiple chains. Specifically, the lattice can be formed by orthogonally combining the chains or graphs with distinct hopping parameters. Fig 2(c) illustrates the two-dimensional (2D) example by combining two different directed chains with hopping parameters of $t_1$ and $t_2$, analogous to a torus formed by joining the edges of a 2D lattice from bottom to top and left to right. The chains in the horizontal or vertical direction can be replaced by directed graphs, as shown in Fig. 2(c) left. Since they are orthogonal to each other, the wave propagating along horizontal and vertical direction do not interfere with each other. Fig. 2(d) shows the results for a 2D case, of which a bright corner state is formed around point G. There are also local minima at the node located at $x=0$ $(y=0)$. The localized states decay along $+y$ $(-y)$ or $+x$ ( $-x$ ) directions with decay constant $t_{\chi}^{-\sum_{i=1}^{p} M_{\chi_i}/\sum_{i=1}^{p}(N_{\chi_i}+M_{\chi_i})}$ ($t_{\chi}^{-\sum_{i=1}^{p} N_{\chi_i}/\sum_{i=1}^{p} N_{\chi_i}+M_{\chi_i}}$), where $\chi = y$ or $x$ and $p = 2$ in this case, indicating independent control of the decay profile along two directions. Edge state can also be formed by applying periodic boundary conditions along one direction, e.g., $x$ direction (Fig. 2(e)) [29]. Such configuration yields perfect edge state along the $y$ direction.

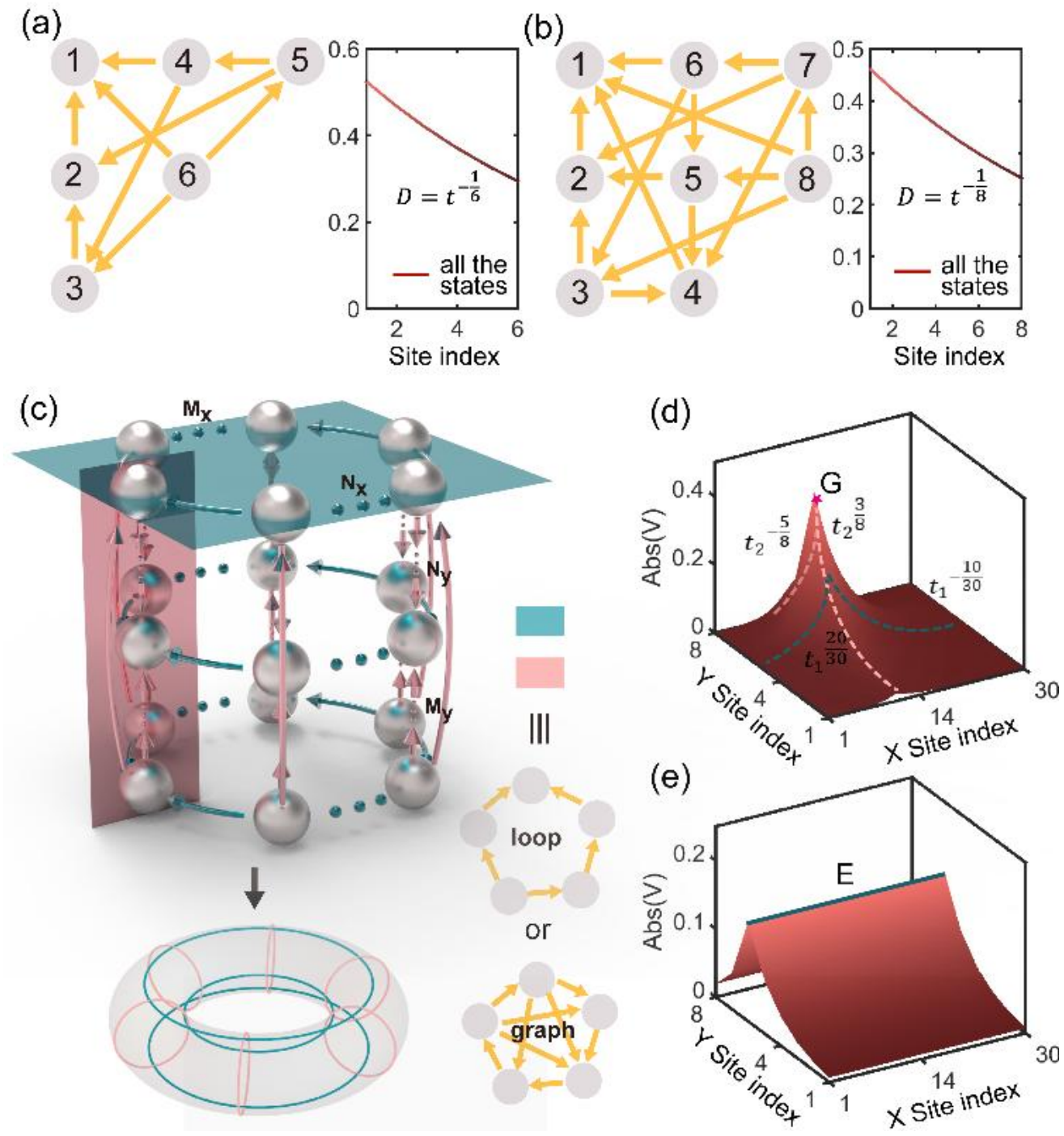

FIG. 2. **Directed graph and higher dimension extension of pure decay mode**. (a, b) Graph configurations that support pure decay mode with (a) 6 sites that each connects with 3 sites and (b) 8 sites that each connects with 4 sites. The corresponding amplitude distributions are shown on the right. (c) General 2-dimension extension of pure decay mode. The red and blue color represent the hopping in 2 orthogonal directions that could be any directed graph or loop discussed in 1D. (d-e) Amplitude distribution when $N_x$=10, $M_x=20, N_y=5, M_y=3$ along each site (d) and when $N_x$=30, $M_x=0, N_y=5, M_y=3$ along each site (e).

Remarkably, the pure decay modes enable the definition of quantized decay charge for each node. Here, the charge for node $\alpha$, denoted as $Q_{\alpha}$, is defined by summing up all the decay constants on the directed edges from that node, i.e., $Q_{\alpha} = \sum_j \log_t(|\psi_{\alpha}|/|\psi_j|)$ where $\psi_j$ represent the wave function of site $j$, and $j$ runs over all the nodes connected to node $\alpha$. In the 1D directed ring shown in Fig. 3a, nodes 7 and 22 carry a charge of $+1$ , node 1 and 15 a charge of $-1$, while the remaining nodes carry a charge of 0. In a 2D example, shown in Fig. 3b, charges of $\pm 2$, $\pm 1$, and 0 can be found. More generally, the charge can also be assigned to directed graphs. For systems with odd numbers of nodes, all the nodes carry integer charges, with one example shown in Fig. 3(c), where the charge for the graph is $[2\ 1\ 0\ 0\ 0\ -1\ -2]$. On the other hand, for systems with even number of nodes, the charges can take half integer values. Fig. 3(d) illustrates the result for a configuration with 8 nodes, where the charges are $[2.5\ 1.5\ 0.5\ 0.5\ -0.5\ -0.5\ -1.5\ -2.5]$. Interestingly, it can be shown that the charge on each node equals to the number of outgoing edges (OE) subtracts the number of incoming edges (IE) divided by 2, i.e. $Q = \frac{Num(OE)-Num(IE)}{2}$, which provide the intrinsic relation $\sum_{\alpha=1}^{N} Q_{\alpha} = 0$ . Therefore, the quantized decay charges always contain balanced positive and negative ones, which are topologically protected by the system geometry.

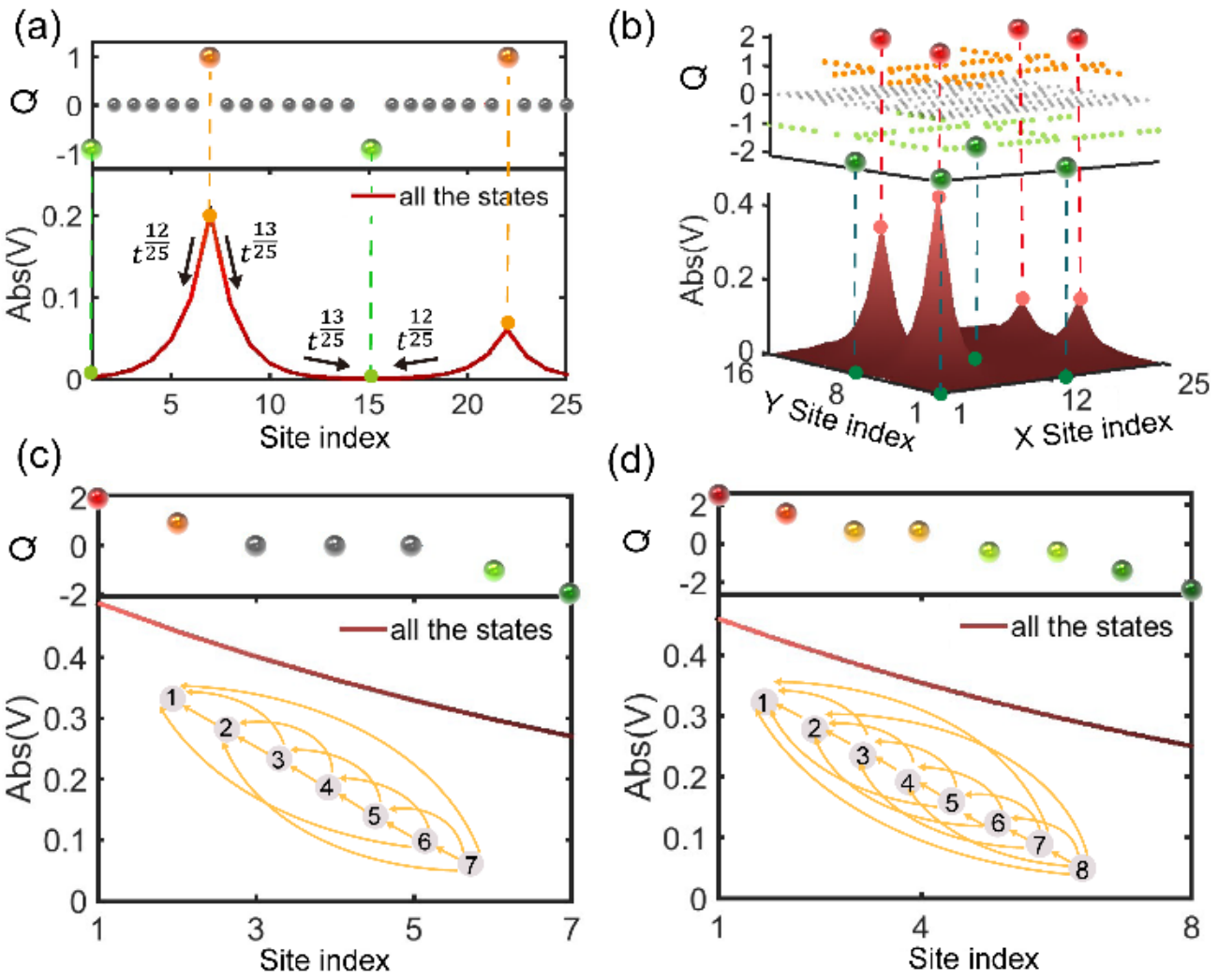

FIG. 3. **Existence of Quantized decay charge in directed system.** (a) Charge distribution for a directed ring with $[N_1 = 6, N_2 = 7, M_1 = 8, M_2 = 4]$. The orange and light green balls represent $+1$ charge and $-1$ charge which are on the localized and delocalized spots. The gray balls represent the charge of 0. (b) Charge distribution for a 2D configuration with $[N_{x1} = 6, N_{x2} = 7, M_{x1} = 8, M_{x2} = 4]$ and $[N_{y1} = 5, N_{y2} = 2, M_{y1} = 3, M_{y2} = 6]$ . The red and green balls represent the charge of $+2$ and $-2$ while others are the same with (a). (c-d) Directed graphs that have quantized charge distribution.

The pure decay skin modes can be experimentally realized at microwave frequencies by designing rings or graphs composed of microwave resonators and amplifiers. We begin with a directed ring consisting of 12 sites, as shown in Fig. 4a. The sample is fabricated by printing copper cladding on a Teflon high-frequency printed circuit board. Each H-shaped resonator supports a dipolar mode corresponding to a site in the tight binding model. Non-reciprocal coupling is achieved using a low-noise amplifier connected between the resonator tentacles, while the reciprocal coupling is provided by a capacitor. The circuit element parameters are derived using an equivalent circuit model of a two-resonator system, based on Kirchhoff's current law (KCL) equations [29]. We then obtain the transmission spectrum by injecting power into the resonators via the Sub-Miniature-A (SMA) port and measure the spectrum at each resonator through additional SMA ports.

Fig. 4b presents the experimentally measured field intensity (green dots) for the 12-site directed ring, with its schematic shown in the inset. The hopping direction between the 12$^{th}$ and 1$^{st}$ site is opposite to that of all other hoppings. In the measurement, the excitation signal is applied at node 1 and the detector is used to probe the voltages at all sites (results for excitation at other nodes are detailed in Supplemental Material [29]). Notably, although the system supports 12 eigenmodes, only the one with the lowest loss (the eigen state with the smallest imaginary part) can be excited. The measured mode profile is in good agreement with the theoretical one (red curve), calculated using coupled mode theory with retrieved parameters. By taking logarithm to both the experimental and theoretical result, we obtain a linear line, as shown in lower part of Fig. 4b, confirming the excitation of a pure exponential mode. A more complex graph configuration with four nodes is also explored, with connectivity illustrated in the inset of Fig. 4c. Each node is connected to all other three nodes, with node 1 acting as the source and the node 4 as the drain. The corresponding Hamiltonian of the system aligns with Eqn. 9, satisfying the conditions for pure decay modes. The measured mode profile and its logarithmic plot, shown in Fig. 4c, exhibit again a pure exponential decay that closely matches theoretical calculations.

The charges for both the directed ring and graph can be calculated from the experimental data. For the ring structure, the charge on the 12th site is $Q_{12} = \sum_{j=1,11} \log_t(|\psi_{12}|/|\psi_j|) = 1.012$, which is very close to the theoretical value of 1. Similarly, for the 4-site graph, the measured charges for the sites are $[-1.51 - 0.514\ 0.492\ 1.52]$ respectively, which also consistent with the theoretical values ($[-1.5 - 0.5\ 0.5\ 1.5]$).

For comparison, we also design a sample consisting of a chain of resonators with OBC, as depicted in Fig. 4d. The system is excited at different sites (site 1, 4 and 6), with the corresponding voltage profiles shown in red, green and yellow, respectively. Despite some variations in the mode profiles, all exhibit oscillatory behaviour, in stark contrast to the non-oscillatory pure decay modes in the above ring and graph configurations.

We have proposed a directed graph framework that enabling the realization of pure decay modes which exhibit quantized decay charges. Based on tailored non-reciprocal hopping, such framework is versatile, applicable to diverse configurations such as directed rings, graphs, and higher-dimensional lattices, as validated through calculations and microwave resonator experiments. The pure decay modes follow a smooth exponential form, enabling monotonic localization of energy and precise control over decay rates. Their amplitude distribution eliminates oscillatory scattering, resulting in greater efficient and robustness against defects. These properties make them suitable for potential applications such as directional energy funneling, directional amplification and information transfer. Such system can be implemented in the realm of acoustics, electronics, and the optical domain using innovative designs, such as nonlinear optics and exciton polaritons, to achieve non-reciprocal coupling [30-33].

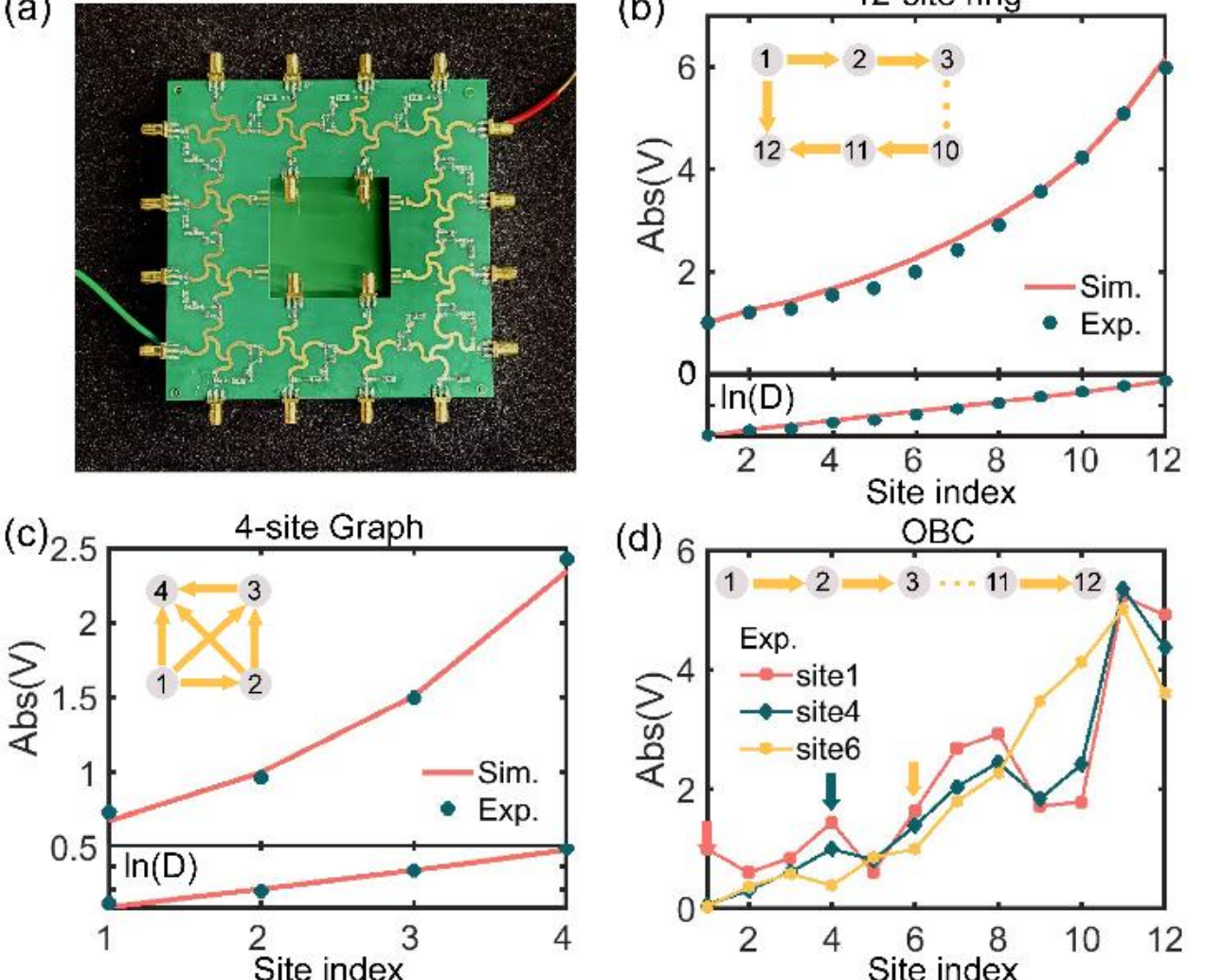

FIG. 4. **Observation of the pure decay mode at microwave frequencies.** (a) Photograph of a sample consisting of 12 identical coupled microwave resonators. (b-c) Upper: experimentally measured (solid circle) and numerically calculated (curves) field intensity distribution of the 12-site directed ring (b) and 4-site for graph (c)_as shown in the inset. Lower: the data (D) after taking logarithm. (d) Measured field intensity distribution for the open boundary 12-site chain (inset) when the source is located at the sites 1, site 4, and site 6, respectively.

We acknowledge useful conversations with Wenan Zang and Kebo Zeng. The work at The University of Hong Kong was sponsored by the New Cornerstone Science Foundation, the Research Grants Council of Hong Kong (AoE/P-502/20, STG3/E-704/23-N, 17309021), the Guangdong Provincial Quantum Science Strategic Initiative (GDZX2204004, GDZX2304001). The work at Zhejiang University was sponsored by the Key Research and Development Program of the Ministry of Science and Technology under Grants 2022YFA1405200, No. 2022YFA1404704, and 2022YFA1404900, the National Natural Science Foundation of China (NNSFC) under Grants No.11961141010, No. 62175215, and No. 61975176, the Fundamental Research Funds for the Central Universities (2021FZZX001-19), and the Excellent Young Scientists Fund Program (Overseas) of China.

#These authors contributed equally.

§To whom correspondence should be addressed

shuzhang@hku.hk

† To whom correspondence should be addressed.

yangyihao@zju.edu.cn

* To whom correspondence should be addressed.

bmin@kaist.ac.kr

[1] Y. Ashida, Z. Gong, M. Ueda, Non-Hermitian physics, Adv. Phys. 69, 249-435 (2020).

[2] C. M. Bender, S. Boettcher, Real Spectra in Non-Hermitian Hamiltonians Having PT Symmetry, Phys. Rev. Lett. 80, 5243-5246 (1998).

[3] I. Rotter, A non-Hermitian Hamilton operator and the physics of open quantum systems, J. Phys. A: Math. Theor. 42, 153001 (2009).

[4] V. M. Martinez Alvarez, J. E. Barrios Vargas, L. E. F. Foa Torres, Non-Hermitian robust edge states in one dimension: Anomalous localization and eigenspace condensation at exceptional points, Phys. Rev. B 97, 121401(R) (2018)

[5] L. E. F. Foa Torres, Perspective on topological

states of non-Hermitian lattices J. Phys. Mater. 3, 014002 (2020)
[6] C. M. Bender, D. C. Brody, H. F. Jones, Complex Extension of Quantum Mechanics, Phys. Rev. Lett. 89, 270401 (2002).
[7] D. Braghini, L. G. G. Villani, M. I. N. Rosa, J. R. de F Arruda, Non-Hermitian elastic waveguides with piezoelectric feedback actuation: non-reciprocal bands and skin modes, J. Phys. D: Appl. Phys. 54, 285302 (2021).
[8] H. -K. Lau, A. A. Clerk, Fundamental limits and non-reciprocal approaches in non-Hermitian quantum sensing, Nat. Commun. 9, 4320 (2018).
[9] H. Hodaei, M. A. Miri, M. Heinrich, D. N. Christodoulides, M. Khajavikhan, Parity-time-symmetric microring lasers, Science 346, 975-978 (2014).
[10] M. A. Miri, A. Alu, Exceptional points in optics and photonics, Science 363, 6422 (2019).
[11] S. K. Ozdemir, S. Rotter, F. Nori, L. Yang, Parity-time symmetry and exceptional points in photonics, Nat. Mater. 18, 783-798 (2019).
[12] H. Shen, L. Fu, Quantum Oscillation from In-Gap States and a Non-Hermitian Landau Level Problem, Phys. Rev. Lett. 121, 026403 (2018).
[13] L. Zhang, Y. Yang, Y. Ge, Y.-J. Guan, Q. Chen, Q. Yan, F. Chen, R. Xi, Y. Li, D. Jia, S.-Q. Yuan, H.-X. Sun, H. Chen, B. Zhang, Acoustic non-Hermitian skin effect from twisted winding topology, Nat. Commun. 12, 6297 (2021).
[14] V. Achilleos, G. Theocharis, O. Richoux, V. Pagneux, Non-Hermitian acoustic metamaterials: Role of exceptional points in sound absorption, Phys. Rev. B 95, 144303 (2017).
[15] C. E. Rüter, K.G. Makris, R. El-Ganainy, D. N. Christodoulides, M. Segev, D. Kip, Observation of parity–time symmetry in optics, Nat. Phys. 6, 192-195 (2010).
[16] H. Zhou, C. Peng, Y. Yoon, C. W. Hsu, K. A. Nelson, L. Fu, J. D. Joannopoulos, M. Soljačić, B. Zhen, Observation of bulk Fermi arc and polarization half charge from paired exceptional points, Science 359, 1009-1012 (2018).
[17] J. Cheng, X. Zhang, M. -H. Lu, Y. -F. Chen, Competition between band topology and non-Hermiticity, Phys. Rev. B 105, 094103 (2022).
[18] T. Dai, Y. Ao, J. Mao, Y. Yang, Y. Zheng, C. Zhai, Y. Li, J. Yuan, B. Tang, Z. Li, J. Luo, W. Wang, X. Hu, Q. Gong, J. Wang, Non-Hermitian topological phase transitions controlled by nonlinearity, Nat. Phys. 20, 101–108 (2023).
[19] T. E. Lee, Anomalous Edge State in a Non-Hermitian Lattice, Phys. Rev. Lett. 116, 133903 (2016).
[20] L. Li, C.H. Lee, S. Mu, J. Gong, Critical non-Hermitian skin effect, Nat. Commun. 11, 5491 (2020).
[21] N. Okuma, K. Kawabata, K. Shiozaki, M. Sato, Topological Origin of Non-Hermitian Skin Effects, Phys. Rev. Lett. 124, 086801 (2020).
[22] K. Zhang, Z. Yang, C. Fang, Correspondence between Winding Numbers and Skin Modes in Non-Hermitian Systems, Phys. Rev. Lett. 125, 126402 (2020).
[23] X. Zhang, T. Zhang, M. -H. Lu, Y. -F. Chen, A review on non-Hermitian skin effect, Adv. Phys.-X 7, 2109431 (2022).
[24] S. Longhi, D. Gatti, G. Valle, Robust light transport in non-Hermitian photonic lattices. Sci. Rep. 5, 13376 (2015).
[25] A. McDonald, A. A. Clerk, Exponentially-enhanced quantum sensing with non-Hermitian lattice dynamics, Nat. Commun. 11, 5382 (2020).
[26] W.-T. Xue, Y.-M. Hu, F. Song, Z. Wang, Non-Hermitian Edge Burst, Phys. Rev. Lett. 128, 120401 (2022).
[27] J. C. Budich, E. J. Bergholtz, Non-Hermitian Topological Sensors, Phys. Rev. Lett. 125, 180403 (2020).
[28] L. Feng, R. El-Ganainy, L. Ge, Non-Hermitian photonics based on parity–time symmetry, Nat. Photonics 11, 752-762 (2017).
[29] See Supplemental Material at *https://doi.org/10.1103/dqf4-6fg5* for more details on the theory and derivation of our non-Hermitian directed graph network.
[30] H. Sahin, M. B. A. Jalil, C. H. Lee, Topolectrical circuits—Recent experimental advances and developments, APL Electron. Devices 1, 021503 (2025)
[31] S. Mandal, R. Banerjee, Elena. A. Ostrovskaya, T. C. H. Liew, Nonreciprocal Transport of Exciton Polaritons in a Non-Hermitian Chain, Phys. Rev. Lett. 125, 123902 (2020)
[32] S. Mandal, R. Banerjee, T. C. H. Liew, From the Topological Spin-Hall Effect to the Non-Hermitian Skin Effect in an Elliptical Micropillar Chain, ACS Photonics 9, 527-539 (2022)
[33] S. Pontula, S. Vaidya, C. Roques-Carmes, S. Z. Uddin, M. Soljačić, Y. Salamin, Non-reciprocal frequency conversion in a non-Hermitian multimode nonlinear system, Nat. Commun. 16, 7544 (2025).
[34] K. Bai, T. R. Liu, L. Fang, J. Z. Li, C. Lin, D. Wan, M. Xiao, Observation of Nonlinear Exceptional Points with a Complete Basis in Dynamics, Phys. Rev. Lett. 132, 073802 (2024).
[35] J. Y. Wu, R. C. Shen, L. Zhang, F. J. Chen, B.B. Wang, H.S. Chen, Y.H. Yang, H.R. Xue, Nonlinearity-induced reversal of electromagnetic non-Hermitian skin effect, https://arxiv.org/pdf/2505.09179 (2025).
[36] H. A. Haus. Waves and Fields in Optoelectronics (Pretice Hall, 1984).

# End Matter

*Appendix A*: *Details of the experimental samples*—Figure S5(a) shows the simulated field distribution of a *H*-shape resonator unit that supports dipolar eigenmode. Based on the unit, we built different samples as shown in Fig. 5 (b-d). The experimental samples are fabricated by printing copper cladding on a Teflon high-frequency printed circuit board, corresponding to the three non-Hermitian configurations studied in the main text: a composite ring, an open boundary chain and a 4-site graph, respectively. As shown in Fig. 5(e), the basic building block for these configurations is a two-resonator element, which is equivalent to the resonant circuit model shown in Fig. 5(f) (the sections framed by dashed lines in different colors represent equivalent parts). In such setup, a microwave signal from source couple to resonator 1 via a lumped capacitor $C_4$, and then inject into resonator 2 through a non-reciprocal coupling circuit, finally coupled to the detector through capacitor $C_4$. The inter-resonator coupling incorporates two distinct mechanisms: a reciprocal coupling $\kappa_1$ via capacitor $C_3$, and a unidirectional nonreciprocal coupling ($\tilde{\kappa}_2$) implemented through an RF amplifier circuit that incorporates a low-noise amplifier (LNA) and a dedicated biasing structure. Key circuit functions include signal-transmission capacitor $C_2$ $C_3$ $C_4$, RF-isolating inductors $L1$ for DC bias passage, and noise-suppressing $R1$ and $C1$ filtering, with complete parameters specified as $R_1 = 2.2\Omega, L_1 = 22\text{NH}, C_1 = 1\text{NF}, C_2 = 0.1\text{pF}, C_3 = 0.2\text{pF}, C_4 = 0.3\text{pF}$.

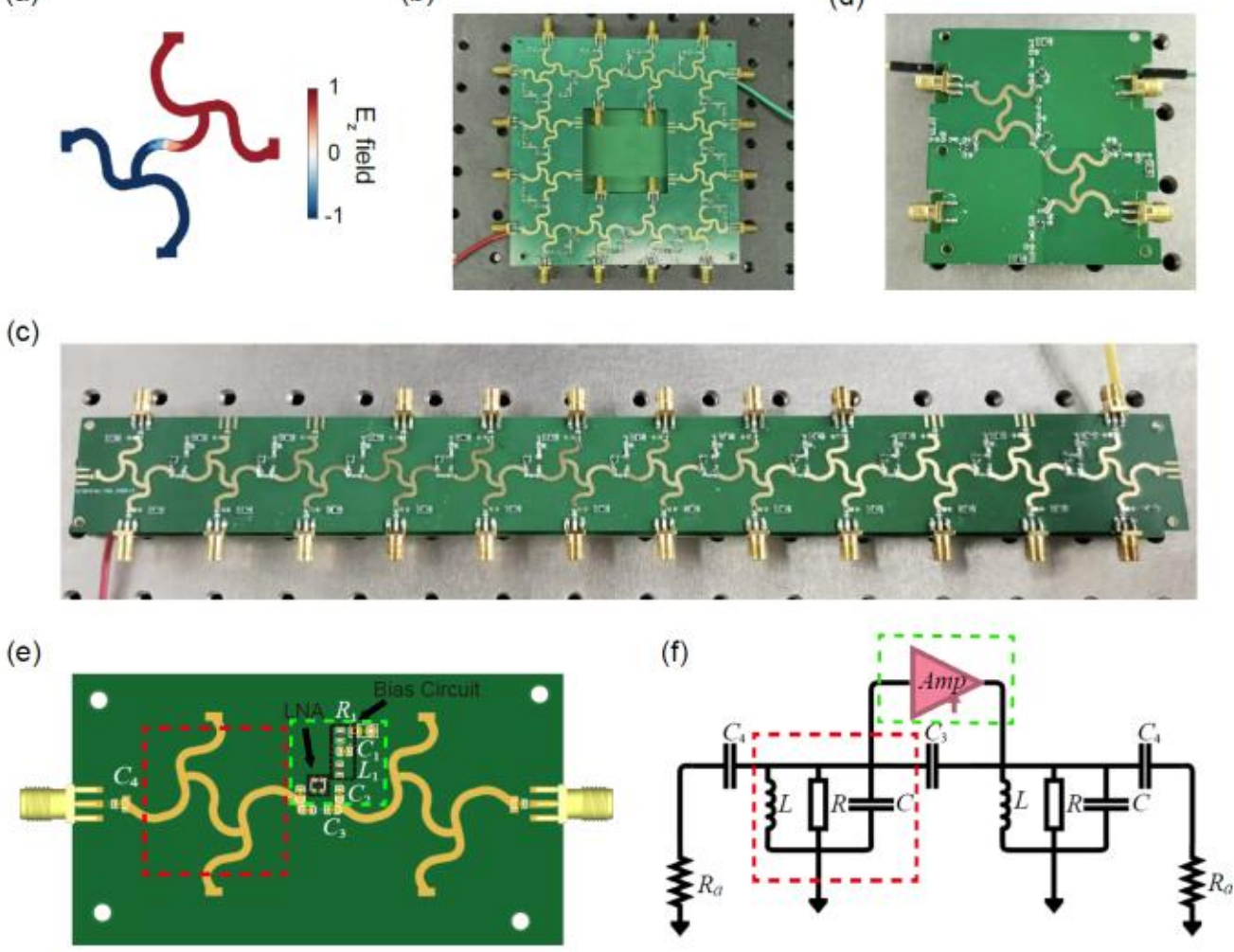

FIG. 5**. Experimental samples for different configurations**. (a) Simulated dipolar eigenmode of a resonator unit. (b-d) Photograph of experimental samples of 12-site ring resonator (b), 4-site graph network (c) and 12-site open chain (d). (e) Detailed circuit elements for the two-resonator system. (f) Equivalent circuit model. The two identical RLC resonators are connected by an amplifier and a capacitor, providing nonreciprocal and reciprocal coupling, respectively. Source and detector (each featuring an internal resistance $R_a$) are coupled to the resonators via coupling capacitors $C_4$.

*Appendix B*: *Derivation of the non-Hermitian Hamiltonian in electric circuits*—In this section, we derive the Hamiltonian of the two-resonator system based on its equivalent circuit model, as depicted in Fig. 5(f) [13, 34]. For operating frequencies near the resonance frequency $\omega_0$, the Kirchhoff's current law (KCL) equations are expressed in a simplified matrix form:

$$\begin{bmatrix} \omega_0 - \omega - i\frac{R_a\omega_0^2C_4^2}{2C_{eq}} - i\frac{1}{2RC_{eq}} & \frac{\omega_0 C_3}{2C_{eq}} \\ \frac{\omega_0 C_3}{2C_{eq}} - i\frac{Y_{21}}{2C_{eq}} & \omega_0 - \omega - i\frac{R_a\omega_0^2C_4^2}{2C_{eq}} - \frac{i}{2RC_{eq}} \end{bmatrix} \begin{bmatrix} V_A \\ V_B \end{bmatrix} = 0 \tag{8}$$

where $C_{eq} = C_4 + C_3 + C$. By taking $\omega_0 = \frac{1}{\sqrt{LC_{eq}}}, \gamma_1 = \frac{R_a\omega_0^2C_4^2}{2C_{eq}}, \gamma_0 = \frac{1}{2RC_{eq}}, \kappa_1 = \frac{\omega_0 C_3}{2C_{eq}}, \tilde{\kappa}_2 = -i\frac{Y_{21}}{2C_{eq}}$, the matrix can be written as

$$\begin{bmatrix} \omega_0 - \omega - i\gamma_1 - i\gamma_0 & \kappa_1 \\ \kappa_1 + \tilde{\kappa}_2 & \omega_0 - \omega - i\gamma_1 - i\gamma_0 \end{bmatrix} \begin{bmatrix} V_A \\ V_B \end{bmatrix} = 0. \tag{9}$$

Therefore, the system's dynamics are directly described by the eigen problem of the Hamiltonian

$$H = \begin{bmatrix} \omega_0 - i\gamma_1 - i\gamma_0 & \kappa_1 \\ \kappa_1 + \tilde{\kappa}_2 & \omega_0 - i\gamma_1 - i\gamma_0 \end{bmatrix} \tag{10}$$

*Appendix C*: *Retrieving the system parameters from coupled mode theory (CMT)*—The transmission spectra for the two-resonator system could be obtained with two channels connected to the SMA port. The equivalent model of the two-resonator system is illustrated in Fig. 6. According to the Coupled-mode theory (CMT) [35, 36], when the wave is incident to the resonator 1, the dynamic equation can be described as

$$\frac{d}{dt}\begin{bmatrix} a_1 \\ a_2 \end{bmatrix} = \begin{bmatrix} -i\omega_0 - \gamma_1 - \gamma_0 & -i\kappa_1 \\ -i(\kappa_1 + \tilde{\kappa}_2) & -i\omega_0 - \gamma_1 - \gamma_0 \end{bmatrix}\begin{bmatrix} a_1 \\ a_2 \end{bmatrix} + \begin{bmatrix} \sqrt{2\gamma_1} \\ 0 \end{bmatrix} s_{1+} \tag{11}$$

where $a_1$ and $a_2$ is the mode in resonator 1(2), $s_{1+}$ represents the incident wave from resonator 1. The transmission coefficient is then

$$s_{21} = -\frac{i2\gamma_1(\kappa_1 + \tilde{\kappa}_2)}{[i(\omega_0 - \omega) - \gamma_1 - \gamma_0]^2 + \gamma_1(\kappa_1 + \tilde{\kappa}_2)} \tag{12}$$

Similarly, when the wave is incident from resonator 2, we could get

$$s_{12} = -\frac{i2\gamma_1(\kappa_1)}{[i(\omega_0 - \omega) - \gamma_1 - \gamma_0]^2 + \gamma_1(\kappa_1 + \tilde{\kappa}_2)} \tag{13}$$

By simultaneously fitting both equations to the measured transmission spectra, we could retrieve the parameters for each circuit.

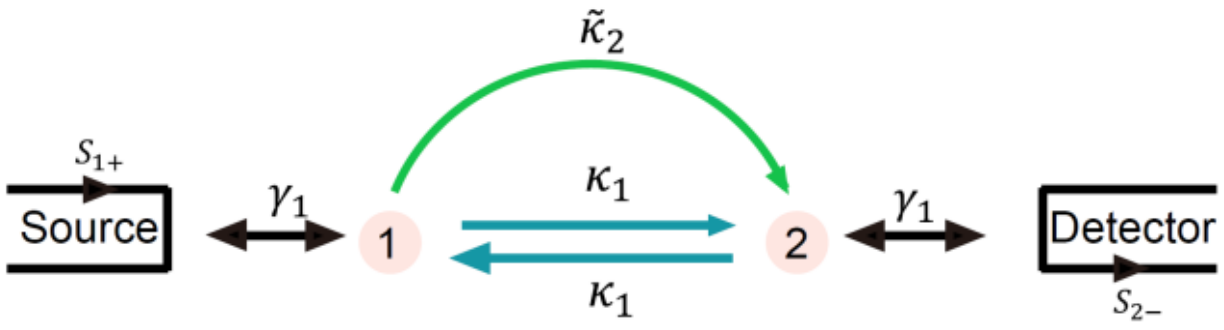

FIG. 6. **Coupled-mode theory (CMT) model of the two-resonator system with nonreciprocal coupling.**

*Appendix D: Experimental result with different source location*—Theoretically, the eigenmode solution suggest the composite ring always supports a perfect pure decay mode. However, experimental insertion losses at the input and output channels introduce minor amplitude perturbations near the excitation sites, which is also verified through the CMT calculation. While the main text presents the pure growing mode amplitude distribution when excited at site 1, Fig. 7(a) shows comparable results for site 4 excitation. Notably, despite localized amplitude reductions, the exponential growth coefficient remains invariant across excitation sites, consistent with CMT predictions. The energy spectrum, derived from experimentally retrieved parameters, is depicted in Fig. 7(b), illustrating all 12 complex eigenmodes. In the experiment, we could excite the mode with the lowest loss, highlighted within the blue rectangle in Fig. 7(b).

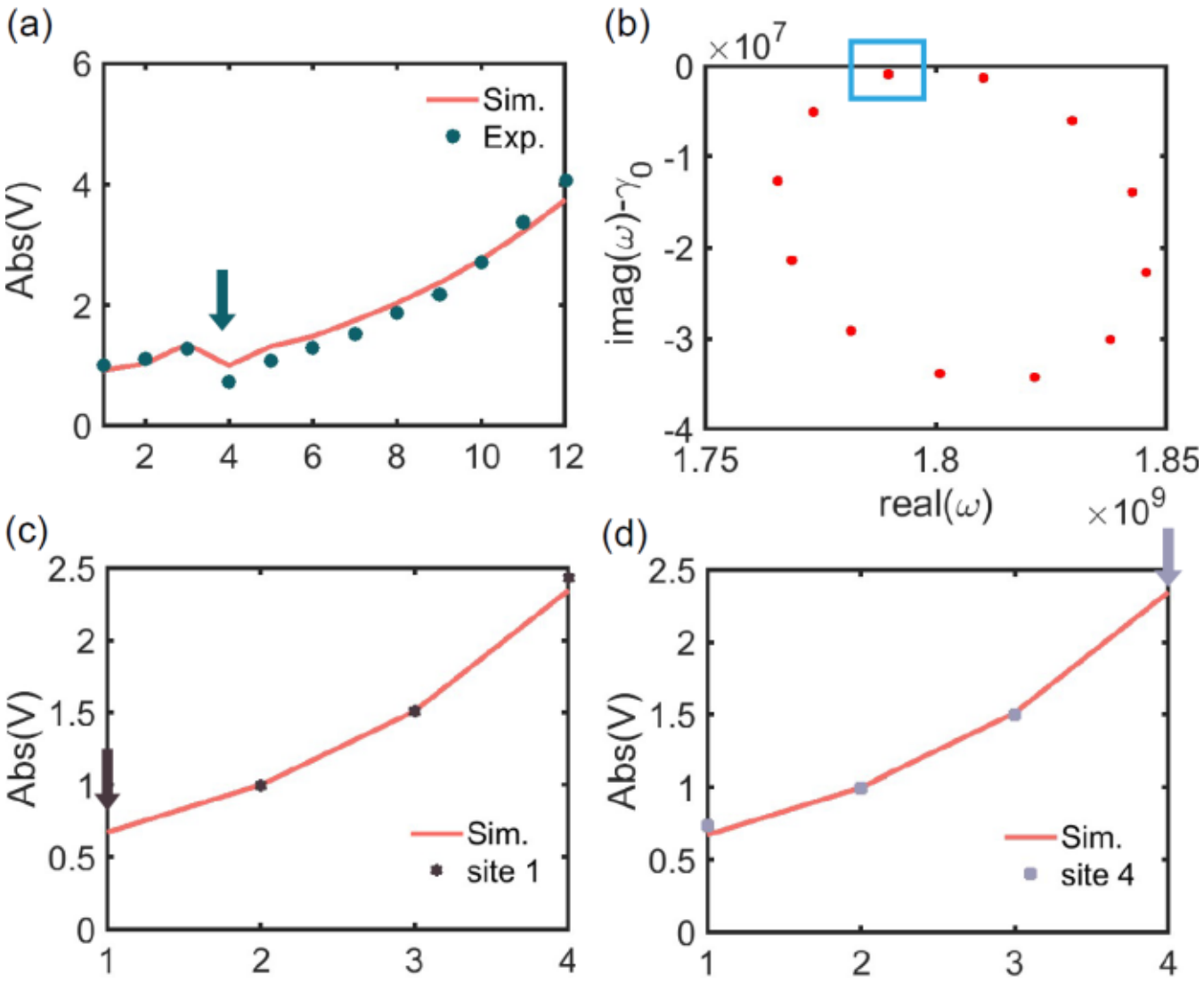

FIG. 7. **Measured result with different source location**. (a) Measured and calculated result of the 12-site ring when the source is located at site 4. (b) Complex eigenmode of the 12-site directed ring. (c-d) Measured and calculated result of the 4-site graph when the source is located at site 1 and 4, respectively. The signal cannot be detected at the input site due to the limited number of channels in 'H' shape resonator.

For the 4-site graph, experimental constraints—specifically, the occupation of a single site by three ports—prevent simultaneous excitation and detection. Consequently, we conducted separate excitation experiments at sites 1 and 4. The field from these experiments were normalized and summed to generate the results presented in the main text, with individual excitation cases shown in Fig. 7(c–d). Crucially, the decay rates remain consistent across different excitation sites, confirming the site-invariant nature of the decay mode.